\documentclass[ajp,pre,groupedaddress,amsmath,showpacs, twocolumn]{revtex4} 

\bibpunct{}{}{,}{s}{}{}

\usepackage{dcolumn} \usepackage{graphicx} \usepackage{natbib}

\begin{document}

\title{Becoming a Physicist: The Roles of Research, Mindsets, and Milestones in upper-division student perceptions} 
\date{\today}

\author{Paul W. Irving} 
\affiliation{CREATE for STEM Institute, Michigan State University, East Lansing, MI 48824, USA}
\affiliation{Department of Physics and Astronomy, Michigan State University, East Lansing, MI 48824, USA}
\author{Eleanor C. Sayre} 
\affiliation{Department of Physics, Kansas State University, Manhattan, Kansas 66506}

\begin{abstract}As part of a longitudinal study into identity development in upper-level physics students, we used a phenomenographic research method to examine students' perceptions of what it means to be a physicist. Analysis revealed six different categories of perception of what it means to be a physicist.  We found the following themes: research and its association with being a physicist; differences in mindset; and exclusivity of accomplishments. The paper highlights how these perceptions relate to two communities of practice that the students are members of, and highlights the importance of undergraduate research for students to transition from the physics undergraduate community of practice to the community of practicing physicists. \end{abstract}


\maketitle


\section{Introduction}

The growth of physics as a field lags behind the growth of all other STEM fields.\cite{NSB2006, NSB2008} The retention of students who have already made the choice to become a physicist is also low\cite{Seymour1997, Hazari2005}. Addressing this retention problem would substantially solve the lackluster growth rate for physics. The development of a professional identity is a fundamental part of student development\cite{Luehmann2007}; the development of an appropriate subject-specific identity is a strong influence on retention of students in a discipline\cite{Pierrakos2009}. However, becoming a professional physicist and identifying oneself as belonging to the physics community is a complicated process that involves overcoming multiple barriers.

Students' progress in their development of a physics identity can also influence their persistence in the study of physics\cite{Shanahan2007}.  There is a strong link between the level of identification with being a physicist and whether or not a student choses a physical science career\cite{Barton2000, Chinn2002, Cleaves2005, Shanahan2007} 
Previous research on identity development in physics has focused on gender differences or on the lack of physics majors of color\cite{Hazari2010, Basu2008, Buck2005}. Recently, this focus has shifted to look specifically at how a student transforms from a student of physics to being a physicist\cite{Irving2014}, broadening the perspective from under-represented groups to physics majors in general. This identity formation process is important to understand\cite{Luehmann2007}. The development of students' identities will help students cope with the continuous change and uncertainty that they will face in life in the 21st century\cite{Brophy2009, Flum2006, Harrell-Levy2010}. It will also affect their interactions within the community of practicing physicists. 

However, different students have different perceptions of what it means to be a physicist\cite{Irving2013}, and for this reason they percieve different experiences with physics as being the end point to their identity formation as a physicist \cite{Irving2013}. This natural variation in the population of physics majors injects an element of uncertainty into research on identity formation. These varying perceptions could influence how students approach and reflect on their development as physicists.  They may also influence the students' efforts to become more central members of the community of practicing physicists\cite{Wenger1998}. The influences may be both explicit and implicit in the students' minds and discourses. If misaligned perceptions among physics students and physicists persist until the end of the students' undergraduate careers, those students may be inadequately prepared for professional life in physics after graduation. The lack of preparedness may be especially acute for students pursuing a career path focused on becoming a central member of the community of practicing physicists, such as through graduate school in physics.

In this study, we conducted a phenomenographic analysis of interviews with upper-division physics students to categorize their perceptions of what it means to be a physicist.  We have two research questions: what do students think being a physicist means? how do their perceptions change over time?

In Section \ref{sec:theory}, we discuss three aspects of our theoretical framework: identity (\ref{sec:ontologies}), Communities of Practice (\ref{sec:cop}), and perceptions (\ref{sec:perceptions}).  In Section \ref{sec:methods}, we outline our methods of data collection and analysis. Sections \ref{sec:categories} and \ref{sec:change} present the  categories of perceptions and how students developed over time, respectively. Sections \ref{sec:catcop} and \ref{sec:changecop} tie our categorizations back to our theoretical frameworks.  We finish with discussion (Section \ref{sec:discussion}) and conclusion (Section \ref{sec:conclusion}).

\section{Theoretical Framework\label{sec:theory}}
\subsection{Ontologies for Identity\label{sec:ontologies}} 

The literature on identity development posits two different ontologies for identity: \textit{identity-as-property}\cite{Gee2000} and \textit{identity-as-activity}\cite{Lave1991}. 

In the property ontology, identity is a thing that people have. It develops over time, can be recognized by oneself or others, and tells you what ``kind of person'' you are\cite{Gee2000}. Research using this ontology has identified many factors that affect the formation of a student's identity within physics. A natural methodology for identifying aspects of identity in this methodology is to conduct interviews or surveys with physics students. The works of Gee,\cite{Gee2000} Carlone (and collaborators),\cite{Carlone2007} and Hazari (and collaborators)\cite{Hazari2010} sit within this strand.

Within Carlone's identity framework\cite{Carlone2007} (as expanded by Hazari\cite{Hazari2010}), the four primary components of influence on a student's identity development are: 
\begin{itemize}
\item interest (personal desire to learn/understand more physics and voluntary activities in this area);
\item  competence (belief in ability to understand physics content);
\item  performance (belief in their ability to perform required physics tasks); and 
\item recognition (being recognized by others as a physics person).
\end{itemize}
Subsequently Potvin and Hazari\cite{Potvin2014} found that interest and recognition to be the main influence on a student self-identifying as a physics person. They also indicated that they believe that identity is a ``quasi-trait'' which can change over time as a result of new experiences. Potvin and Hazari also found that the four components of interest, competence, performance, and recognition are ``contingent on students perceptions of what physics is.'' The data from which Hazari\cite{Hazari2010} developed her framework from was mainly high school students.  It focuses on students identifying as a ``physics person.'' The study presented in this paper uses the Hazari et el. framework as a guide.  However, our context is upper-division physics. By the time students take classes in upper-division physics, their identities might have transitioned. They could have moved from an identity as a physics person to an identity as a physics student (a subject-specific identity in physics). There is also the possibility that they are on the verge of transitioning again from having a subject-specific identity as a physics student to an identity as a professional physicist.

Discussing transitions of identity within a discipline introduces us to the alternative ontology of ``activity.'' In the activity ontology, research focuses on the practices of identification within a discipline\cite{Jawitz2009}. In this ontology, we see identity in the activities that participants perform, and the ways participants position themselves with respect to the legitimate practices of a discipline \cite{Lave1991}. Research using this ontology has covered wildly diverse communities, from insurance agents\cite{Lave1991} to engineering students\cite{Stevens2005}. Within physics programs, research has examined groups from Hispanic students in introductory physics\cite{Goertzen2013} to physics majors in advanced laboratory\cite{Irving2014}. Research conducted from this perspective typically employs the communities of practice framework and is discussed in detail in the next section.

Both ontologies for identity are fruitful and interesting for research. They naturally have overlapping domains. For example, a student's interest in the subject of physics could motivate their engagement in more authentic practices of the community of practicing physicists or their membership in more communities that are physics centered. Taken together, the ontologies suggest that an individual's interest and motivation to become a physicist both affect and are affected by her participation with other physics-interested and physics-identified people. The research in this paper examines students' perceptions of the physics community and the field of physics, especially as they relate to the students' positioning of themselves within the field. The reflection on their attitudes and expectations influences their current and prospective roles in their various physics communities.

\subsection{Communities of Practice\label{sec:cop}}

In previous literature on communities of practice, a community of practice is defined by having the following key characteristics:
\begin{itemize}
\item The individuals within a community form a group, either co-located or distributed\cite{Coakes2006} 
\item The formed group has common goals or a shared enterprise\cite{Wenger1998}
\item The group shares and develops knowledge focused on a common practice\cite{Barab2002}. 
\item The group shares mutually defined practices, beliefs, values, and history\cite{Irving2014}
\end{itemize}

Within the communities of practice theoretical framework, learning is conceived as a trajectory towards being a central member of a community\cite{Wenger2000}. This central membership is achieved by engaging in the legitimate peripheral practices of the community while being guided in these practices by central members of the community\cite{Barab2002}. In the past, the community of practice framework has been applied to the undergraduate context\cite{Donath2005}. From this perspective, activities such as taking a class or taking an exam would be perceived as being legitimate peripheral practices of the specific undergraduate community of practice\cite{Irving2014}. Practices are generally considered peripheral as long as they are external to the practices of the central members or if there is still guidance from a central member. It is this guidance that makes it a legitimate practice. But some classes would be considered more central than others, for example, an advanced laboratory class would be considered more central than a introductory physics course.  Applying this framework to the undergraduate physics context then results in taking a quantum mechanics class or doing undergraduate research being interpreted as legitimate peripheral practices of the community of practicing physicists. From this perspective, professors act as the central participants who are guiding peripheral participants on a trajectory to the core of the community of practicing physicists.

But applying the communities of practice framework to the undergraduate physics context is not as straightforward as described in the previous paragraph. A major complication is that students participate in several overlapping communities of practice at the same time\cite{Avery2001}. A simple example would be a physics student who also plays on a sports team, but is also a member of a research group. Membership to both of these groups implies membership in multiple communities of practice. A more complicated example is that being a member of the research group might involve a larger collaboration with another research group that then implies membership of overlapping communities of practice. Given this overlap it is safe to assume that the knowledge and practices being learned in one community affects practices in another\cite{Aschbacher2010}. It can also be true then that when communities of practice have different values, individual members may have difficulty importing practices from one community to another\cite{Aikenhead1996}. 

We have argued before\cite{Irving2014} that undergraduate students maintain membership in two overlapping communities of practice: the undergraduate physics community of practice and the community of practicing physicists. These two communities of practice refer to two different identities. The undergraduate physics community of practice refers to a subject-specific identity\cite{Creighton2007} as opposed to the community of practicing physicists that correspond to an identity as a professional physicist. These two communities overlap in two ways: central participants in each community have some of the same people, and the two communities share some practices.  However, a fully-developed subject-specific identity does not result in full central membership in the community of practicing physicists. Instead, this development is only part of the trajectory to becoming a central member of the community of practicing physicists. The borders and overlaps for these two communities has not been explored in detail and the legitimate peripheral practices of the undergraduate physics community will be dependent on context. For example, Kansas State University does not require undergraduate research as a prerequisite to obtaining a degree, but other institutions may stipulate this requirement. In this context this would make research related practices overlapping between the undergraduate physics community and the community of practicing physicists. 

From this perspective, some of the practices of the two communities are the same; participation in one practice may help a legitimate peripheral participant become more central in both communities. However, some legitimate peripheral practices result in an accelerated trajectory towards central membership of the community of practicing physicists. For example, conducting research or writing a grant proposal are more likely to be more central legitimate practices of the community of practicing physicists, and participation in those activities may help a student become more central as well as help her develop her identity. On the other hand, attending class and doing well on exams are central practices of the community of practice of undergraduate physics students, but are more peripheral practices to the community of practicing physicists. They are more likely to accelerate a student's trajectory towards central membership of the former community than the latter. Many undergraduate physics programs (although perhaps not explicitly) encourage students to engage in more central practices of the community of practicing physicists through participation in undergraduate research. Concurrently, undergraduate research programs can have a major impact on the development of students' identity.\cite{Hunter2006}

Students' participation in these diverse physics-related communities can shift their identities and perceptions of community practices and goals. Through engagement in these multiple communities, students experience different perspectives (attitudes, principles, and expectations\cite{Hazari2010}) that make up a professional physicist identity. As a result of experiences, students can transition from identifying themselves solely as learners-of-physics to identifying themselves as belonging-to-physics\cite{Osterman2000}. 

\subsection{Perceptions\label{sec:perceptions}}


A large body of research explores physicists' and physics students' perceptions of learning physics and physics as a field of academic study. Often this research has focused on students' perceptions of learning physics\cite{Prosser1996, Adams2006} or their perceptions of various teaching practices\cite{Brekelmans2006, Adams2006}. Other studies have focused on students' conceptions of the subject of physics itself at various levels\cite{Barmby2006}. 

In the area of identity development, there is a body of work focused on the perceptions of female students either studying or conducting post-graduate research in physics\cite{Moss-Racusin2012} or allied fields\cite{Potvin2012}. In teacher professional development, teachers' perceptions of their professional identity reflect their personal knowledge of this identity\cite{Beijaard2000}. A teacher's professional identity is a combination of her self-perceptions along three themes: as subject matter experts, as pedagogical experts, and as didactical experts\cite{Beijaard2000}. This combination of self-perceptions could be interpreted as being associated with the different communities of practice of which a person is a member. For example, their self-perception as subject matter experts is related to their position within that subject community of practice.

For this reason, her perceptions of what the practice entails will govern her participation in the practice itself. It follows that physics students' perceptions of the practices of physicists are intricately linked with the practices they engage in within that community. These perceptions are also linked to their perceptions of themselves as (potential) physicists and their perceptions of who physicists are and what they do.

In light of our theoretical framework, we refine the research questions of this paper: how does participation in overlapping communities of practice affect students' perceptions of the central practices of physics and the development of students' physics identities?

\section{Methodology\label{sec:methods}}

The phenomenographic research methodology was introduced in Marton and Saljo's seminal research study examining students approaches to learning\cite{Marton1976a, Marton1976b}. Since then, the phenomenographic methodology has become a widely used methodology for research on learning and teaching\cite{Bowden1992, DallAlba1993, Walsh1993, Ramsden2002, Entwistle1983, Prosser1999, Laurillard2002, Ramsden1993, Olympiou2012, Lee2008}. A phenomenographic study usually focuses on a relatively small number of subjects and identifies a limited number of qualitatively different and logically interrelated ways in which a phenomenon or situation is experienced or perceived.

This idea of qualitatively different ways of experiencing a phenomenon has been validated and reinforced by the theory of variation and awareness\cite{Marton1997, Trigwell1997, Bowden2004, Marton2004, Marton2005}. This theory states that there are a limited number of qualitatively different ways in which something experienced can be understood. The limit is set by the constituent parts or aspects of the experience that are discerned and appear simultaneously in people's awareness. A particular way of experiencing something reflects a simultaneous awareness of particular aspects of the phenomenon. Another way of experiencing it reflects a simultaneous awareness of what aspects (more aspects or fewer aspects) of the same phenomenon are experienced\cite{Marton1997}. For this reason, it is the variation in the way in which critical aspects of a particular phenomenon are discerned that constitutes an individual's experience of that phenomenon\cite{Linder2003}.
 
To transfer the theory of variation to our study, an investigation into students' identity development over time is an examination of the variation in the critical aspects that influence their professional or subject specific identity. Then we re-examine how these critical aspects change over time and if new critical aspects begin to influence the student's identity. The typical outcome of a phenomenographic study is the researchers' interpretation of people's experiences/perceptions in relation to an aspect of the world\cite{Mann2007}. In the case of this study, it is the development of their professional physics identity. The results of a phenomenographic study are typically presented as a set of categories. Our categories (Section \ref{sec:categories}) represent the variation in how this group of students perceives (in this case their attitudes towards and expectations of) physicists. 

\subsection{Data collection}

The primary data for this analysis come from semi-structured interviews with students who were recruited from upper-level physics courses in electromagnetism, classical mechanics, modern physics and advanced laboratory at Kansas State University (KSU). We developed a 45-minute semi-structured interview protocol drawing on identity formation\cite{Hazari2010}, epistemological sophistication\cite{Elby2010}, and metacognition literature\cite{Brown2009a}. Interviews, which were video-taped, began with a discussion of the student's prior history with physics up to the time of the interview and segued into questions about their present physics experiences in class, their attitudes in physics, future career plans and finally a discussion on physicists. Twenty-one students initially chose to participate in the study.  They were all enrolled in upper-level physics classes at the time and ranged from sophomores to seniors. The initial set of interviews was carried out over a two-week period near the end of the spring semester. The sample was comprised of 3 female and 17 male interviewees. 

The second set of interviews was conducted 3-6 semesters later, depending on the availability of the students. There was at least a 3 semester gap because the research was designed to look at students before and after they had completed a upper-division laboratory course with the idea that they would engage in authentic practices of physicists in this environment. The longer gaps (e.g. 6 semesters) are a result of students' busy schedules and our difficulties with getting them to volunteer for interviews until they were ready to participate.

In the second set of interviews, we used a similar 45-minute semi-structured interview to explore similar topics. Significant differences in interview protocols centered on students describing their physics-related experiences between the first interview and the second interview and less time spent on their prior history. Of the 20 students who chose to participate in the first set of interviews, only 7 were available to be reinterviewed. The sample consisted of 7 male interviewees. 

The role of undergraduate research plays an important part in the data analysis. In reference to the context of the study, the KSU physics department encourages students to get involved in a research group, but does not make a substantial coordinated effort to include them. Some faculty welcomes research with undergraduates; others are more guarded and only make exceptions for special cases. Students are not required to conduct research to graduate, and KSU does not require a capstone class or research seminar where students might conduct research.

\subsection{Data Analysis}

For each set of interviews, the responses to the questions were analyzed initially by an individual researcher and the robustness of the categories was tested by a fellow member of the research team. The initial focus on an individual researcher is due to the process of phenomenographic analysis.  The outcomes the process produces are constituted through the relationship between the researcher and the data\cite{Yates2012}. We validate the outcomes of the study through an iterative two-stage peer review process\cite{Cresswell2000}. 

In the individual process, each transcript was repeatedly read, often in one sitting, in order to become acquainted with the transcript set as a whole. For each sitting of the transcript, the focus of awareness was on one particular aspect of the transcript or theme. For example, one theme was how students view the relationship between physics and mathematics.  Each theme should emerge from multiple places in the interviews. In response to a question about when they first became aware of their interest in studying physics, students may describe that they were good at math but wanted to apply it to more real-world situations.  Alternately, the math-and-physics theme might emerge when students describe their experiences in their current classes, or their difficulties distinguishing between math and physics as they progress through the curriculum. 

For each emergent theme, we explored the variation in that theme amongst all of the students' descriptions of their experiences, and attempted to relate each theme to their identity development. In subsequent readings of the transcripts, the focus of awareness shifted to other aspects of the students' discourse (for example, positive affective descriptors).  Not all emergent themes spoke to the students perceptions of being a physicist and so did not inform the discovered categories. Two important emergent themes that informed the categories were students' affective responses to physics, and student's conceptions of when they will consider themselves a physicist. 

After developing themes, the next step was to make a set of notes that recorded all information (including emergent themes) that was perceived to be critical to the students' perceptions of physicists. The analysis then moved to seeking out the critical similarities and differences (i.e. variations) between the notes for each student and each theme. However, the focus was not solely on the notes and instead involved working concurrently with the notes, transcripts, and videos as the notes often lacked the depth of completeness that the videos contained. The next step in the analysis was to examine these critical aspects of students' perceptions of physicists in more detail by examining cases of agreement between students on a particular critical aspect. In this stage of the analysis, the variation between these critical aspects was also identified and explored in detail. For example, for the theme of when a student will consider themselves a physicist, students descriptions of this perception were explored for similarities and differences. But, also, for students who had similarities in their descriptions the critical aspect of this similarity was discerned and compared against the critical aspect of another group of students who had a different shared perception of when they would consider themselves a physicist. 

The cases of variation and agreement of critical aspects of the identified themes were then utilized to form categories of description (an outcome space) of the different perceptions of physicists. This outcome space is preliminary in nature. For each category, the groupings of notes were re-examined to find cases of both agreement and variation within the notes. This process was to ensure that the categories did describe the variations in the perceptions of physicists from the set of students interviewed faithfully and empirically, and to sharpen the differences between the categories. 

The next step in the analysis process was to begin the first part of the peer-review process. This step involved giving the transcripts and preliminary categories of description to another member of the research group who then examined the robustness of the categories individually. Once they had analyzed the categories and transcripts in concert, a discussion would occur where we identified differences in interpretation, and a negotiation process would begin. The negotiation process involved each researcher reviewing the interviews and identified themes in detail in order to provide evidence of their interpretation. This process continued until both researchers had co-developed an interpretation of the data resulting in refined categories describing how students perceive physicists.  Together, we produced descriptions of the categories. The final part of first peer-review process involved taking extracts and statements from the transcripts that would give substance and support to the categories.

The final reliability and validity check was the second peer-review process that involved presenting the research to the physics education group at Kansas State University in order to conduct a peer review\cite{Lincoln1985}. The peer review focused on the group challenging the researchers' assumptions and asking the researchers questions about method and interpretation. After this review, minor changes were made to each category.

We repeated the same process for the second set of interviews, focusing on whether or not students' perceptions changed. The previously identified themes were used to analyze what aspects of their perceptions had changed over time and what reasoning (if any) students gave for this change. This analysis involved examining each student's two separate interviews together as a set.  We identified the important aspects of being a physicist perceived in the first interview and examined how the perception of these aspects changed over time. We conducted a similar analysis process (individual theme identification and then categories of descritpion development followed by peer-review) with transcripts analyzed with a particular focus on one theme at a time to compare and contrast changes in perceptions. 

The completeness of the variation in students' perceptions of being a physicist in the second set of interviews is limited due to the sample size. An important example of this limited completeness is apparent in the sparsely populated categories of perception discovered in the second set of interviews. These missing categories of perception should not be interpreted as perceptions that disappear with time. Instead, we can no longer discover this perception within this set of data with the limited sample size available to us. However, a complete variation in the perceptions of being a physicist is not the focal point of the analysis of the second set of interviews. Our data for examining a change in perception is as complete as possible given this population of students. 

In Section \ref{sec:categories} we present the categories of perception discovered from the first set of data and provide extracts and statements taken from the transcripts that would give substance and support to the categories. In Section \ref{sec:change} we present extracts and statements taken from the second set of transcripts that support our analysis of students' shifts in perception.
 
\section{Categories\label{sec:categories}}

\begin{figure}[tb]
\begin{center}
\includegraphics[width=7cm]{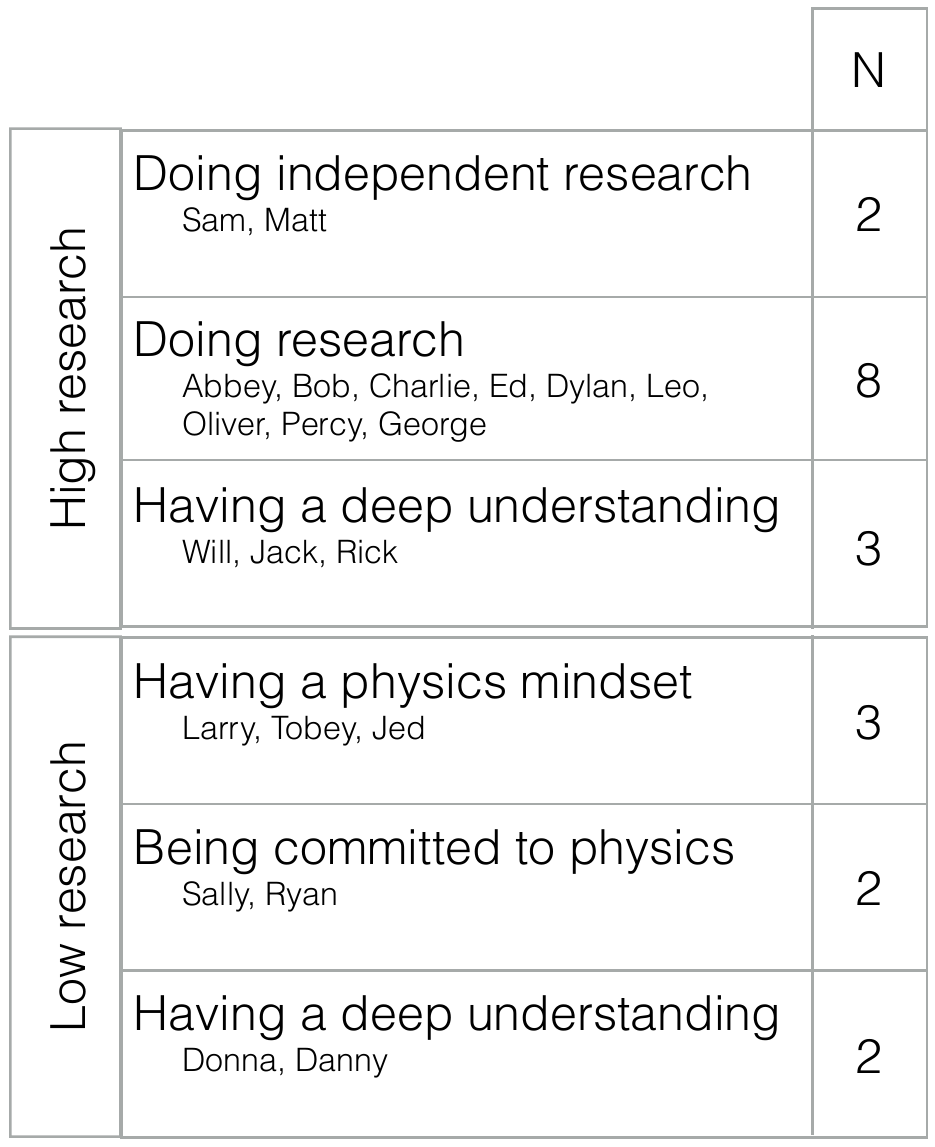}
\caption{Categories and subcategories of students' perceptions, and the number and pseudonyms of students in each.}
\label{fig:categories}
\end{center}
\end{figure}

The phenomenographic analysis of the interview data resulted in six distinct categories of description for students' perceptions of physicists. None of these categories is ``bad'' or ``good''; we present them without value hierarchy. Within each of those broad categories, three subcategories emerged (Figure \ref{fig:categories}.
Briefly, the categories are: 

\begin{description}
\item[High research / Doing Independent Research] Research is very important to being a physicist and when I am doing research by myself, such as when I am a Principal Investigator, I will be a physicist.

\item[High research / Doing Research] Research is very important to being a physicist and when I am doing research I will be a physicist.  The major difference between this category and the preceding one is that students in this category do not emphasize that they must lead the research, only participate in it.

\item[High research/  Deep Understanding] Research is important to being a physicist but so is developing a mastery of the subject.

\item[Low research / Mindset] I am already a physicist because I have the interest and mindset of a physicist.  Doing research is unimportant to me.

\item[Low research / Commitment] I am already a physicist because I have made a commitment to the subject, such as by declaring a physics major or minor.

\item[Low research / Deep Understanding] When I develop a mastery of the subject I will be a physicist, regardless of whether I do research.

\end{description}

A detailed description for each category is presented below with relevant quotes from the students taking part in the interviews presented to illustrate the origin of the categories in the data and to provide a more thorough description of each category.

\subsection{High research / Doing Independent Research}

This category is very similar to the following one, but with the added emphasis on independence in research as a deciding factor of when one becomes a physicist. This is an independence informed by the students previous or current experiences of doing research which is evidenced in both the descriptions students in this category provide of research and the role it plays in being a physicist and the lack of detail in those same descriptions students of the high research category provide. There is also a sense of ownership and positioning themselves as the researcher that is present in the independent students descriptions of their perceptions of research that is absent in the doing research category. The doing research students tend towards generalities and impressions of what research might entail. There are three students who fall into this category. All three students were engaged in undergraduate research at the time of interview and have been for at least one semester. Another commonality between this set of students is that all of them declared only one major, and it is in physics.

In the following extract, Matt discusses whether he considers himself a physicist at this point in his undergraduate career. 

\begin{description}
\item[Int:] Do you consider yourself a physicist?
\item[Matt:] Physicist in training
\item[Int:] How do you think you move from the point you're at now, physicist in training, to being an actual physicist? 
\item[Matt:] Getting to the point where you can design your own experiments and set up research equipment and try and figure out something on your own without having to resort to looking up the answer.
\end{description}

The threshold for Matt to move from a physicist in training to being a physicist is not just doing research but doing research that originates with him. The reference to looking up the answer would also seem to indicate that the research Matt has been conducting to date has involved being able to reference previous work. From Matt's perspective, this might see him interpreting his current research as not novel and so not truly what a physicist does. Matt also places an emphasis on ownership with reference to ``own experiments" and ``figure out something on your own" which is not a position that those in the doing research category make.

Sam expresses a similar threshold for when he will consider himself a physicist:
\begin{description}
\item[Sam:] I feel that when you are, when you truly become a physicist\dots is when you have your own ideas and your own project basically that you're working on not being spurred by someone else telling you to do it and you're asking for money and grants for that, I really feel thats when for sure you become a physicist.
\end{description}

Both Sam and Matt emphasize the need for ownership of the research they are conducting as the point at which they will consider themselves physicists. This emphasis on ownership and independence is an important distinction from High Research/Doing Research category where the pure act of doing research is considered the threshold point of becoming a physicist. Given the similarities in the categories the other important aspects of the High Research/Doing Research independence category are included in the discussion of the next category: High Research/Doing Research.

\subsection{High research / Doing Research}
One of the dominant descriptors of this category of description is that physicists answer the unanswered questions, and they do so by conducting research. The focus on physics as an area of research that can answer questions and offer the opportunity to discover something unique is the motivation for these students to become a physicist. Unlike students in the former category, about half of the students in this category are double majors in physics and cognate fields. 

The following extract is from Abbey who is a double major in physics and engineering:
\begin{description}
\item[Int:] To you what is a physicist?
\item[Abbey:] Hmm, that's a good question, I would say a physicist is someone who is trying to answer some of the unanswered questions, trying to prove the impossible.
\end{description}

A double major in physics and chemistry describes why he currently has a preference for becoming a physicist:
\begin{description}
\item[Ed:] Sometimes it feels with chemistry it's not quite on the cutting edge quite like physics is, there's research, but it's not quite as grandiose I suppose as you might see in physics. 
\end{description}

Romantic descriptors of physics are not exclusive to the high research/doing research category but usually come when students are comparing physics to their other major. The students in the independent research category tended not to discuss physics in this manner but that may be due to them being only physics majors and so don't have another subject to compare to physics. Another possible reason is that they had more authentic experiences with physics and realized the reality does not necessarily compare to the romantic notion. The grandeur and impressiveness that students in this category relate to physics research is also evident in their assertion that one is not a physicist unless they are doing new research or discovering something new about the physical world. 

Percy is a physics major who has not been involved in research as of yet:
\begin{description}
\item[Percy:] I think learning is a necessary thing but otherwise it has to mean something, you have to find something out, find something new, find out something new.
\item[Int:] That would be research then would it?
\item[Percy:] Yeah
\item[Int:] So you have to do some research
\item[Percy:] Yes to be a true physicist
\end{description}

In the above quotes Abby and Percy do demonstrate some agency but the emphasis is on the novelity of the discovery which is a repeated theme for students in this category. Again one of more distinguishing factors between the doing research and doing independent research is what is left unsaid by those in the doing research category. Doing research students do not position themselves as the researchers, do not explicitly identify why doing reseach is an important part of being a physicist, and do not describe in any great detail what doing research entails or how it might be independent. Students in the \textit{High research / Doing Research} category do however, identify the threshold point for when becoming a physicist as when one starts doing research. 

In the following extract, Charlie (a physics major) discusses why he thinks it is important to transition to doing research:
\begin{description}
\item[Int:] So how do you think one moves from potential physicist to actual physicist?
\item[Charlie:] I think one good indicator would be research. Like, I haven't done any research yet. There is a difference between learning something and actually doing it. Someone can learn basketball, like read about basketball, but they are not a basketball player until they play basketball.
\item[Int:] Right
\item[Charlie:] In the same way you can learn about physics, you can read all the stuff, but you're not actually a physicist until you actually [pause] do physics, research physics.
\end{description}

When asked to describe what physics research entails, this group of students is understandably vague when it comes to their descriptions as they have not been involved in research as of yet. 

Leo, who is a double major in engineering and physics, describes research as the following:
\begin{description}
\item[Leo:] I would imagine it's designing experiment type things and then you know, carrying those experiments out and collecting data and then interpreting it I guess [laughs], that would be my idea of what a physicist does.
\end{description}

Leo's description is coherent and not inaccurate. It lacks some of the less-discussed elements of the realities of research, such as becoming a member of a research community by submitting your research for peer review. We don't expect an accurate portrayal of the intricacies of research from an undergraduate unless he has had research experience. However, this element of the category  highlights the different conceptions of research that students have, especially when this description is compared to the previous category's emphasis on independence. This lack of emphasis on independence can by observed by Leo not taking ownership over the research he is describing unlike Matt who consistently displays ownership over research by consistently referencing "my own" when discussing research.

The final important detail of this category that is also present in the independent research category is that all students in it do not perceive themselves to be physicists (yet). They often describe themselves as ``aspiring physicists'' or ``up-and-coming physicists''.  Because they haven't conducted research yet, these students stop short from identifying as physicists.

\begin{description}
\item[Int:] So at this moment in time what would you classify yourself as?
\item[Charlie:] A learning physicist, I could never classify myself as a physicist now, maybe a potential physicist.
\end{description}

\subsection{High research/Deep Understanding}

The \textit{High research/Deep Understanding} category of description also has an emphasis on research as a major part of its description. The students in this category agree that research is important. However, neither Will nor Rick speak in any great detail about what research entails nor why it is important to do it to become a physicist or what it is about research that makes one a physicist. They identify the physicists they know (their professors) as doing research and they intend to obtain a research experience themselves. More importantly to them, they believe they will become physicists when they have mastered a certain amount of physics concepts or obtained a certain amount of knowledge. 

In the following extract, Will (a physics major) describes how he believes he will move towards being a physicist from his current position where he self-identified as an aspiring physicist.
\begin{description}
\item[Int:] How do you move towards being a physicist?
\item[Will:] Just the mastery of, especially the basics like, I just learned magnetism and electricity, just seeing my professor do it on the board and how, how much he knew about it, and how much he knew about other areas of physics\dots the amount of knowledge.
\end{description}

Rick is another physics major who identifies research as being important.  He will feel like a physicist when he has obtained his degree. The following extract indicates that Rick thinks a degree equates to knowing a body of physics knowledge that is enough to become a physicist.
\begin{description}
\item[Int:] What makes someone a physicist in your mind?
\item[Rick:] Knowing enough about, not just knowing about physics theory but also being able to apply it and teach it and I guess the best way to measure that would be getting a degree, that's what they are there for, so get that\dots I guess I'm a physicist when I get my degree.
\end{description}

Neither Rick or Will address a need for research in either of the above extracts but both have been placed in this category because they do make several reference to research throughout their original interviews. For example: Will indicates that he should do research and is seeking a research experience and Rick indicates that "the physicists that I know either teach or do research". Research is still a part of being a physicist but a deep understanding is more important to these students at this time. To the students of this category of description, an attainment of a certain amount of knowledge and understanding must occur in order for one to be considered a physicist. Interestingly, like the previous categories described, none of the students placed in this category would consider themselves physicists at this moment in time as the threshold point for them is something they have not yet achieved.

\subsection{Low research/Mindset}

The first three categories all had the theme of research as an important element. Research inherently adds a degree of exclusivity to these three categories which is not matched in the next three.  The categories of perceptions of physicists that emphasize low research are the most inclusive of the categories. 

Students in the \textit{Low research/Mindset} category believe that anyone who is interested in physics can be a physicist. It is their perception that there is a good chance you are already a physicist if you have thought about the world in a certain way.

Students in this category believe that anyone is a physicist if they engage in any physics practices (not necessarily formal or/and structured) even if this is just reading a book about physics.
\begin{description}
\item[Int:] Why are they a physicist?
\item[Jed:] Why are they a physicist, just liking it really, I mean anyone can be a physicist if they show interest in it, I mean people think you need a lot of schooling to be a physicist, but anyone can be a physicist, anyone can be a scientist really its just whether or not you have that interest in it in my opinion.
\item[Int:] Given that definition you must consider yourself a physicist then?
\item[Jed:] Yes I definitely do
\end{description}
Jed's perspective is in opposition to the majority of the previous categories of description and consistent with the fact that the students in this category already believe themselves to be physicists. 

Another significant departure from the previous descriptions is that you don't have to do research to be a physicist.
\begin{description}
\item[Int:] What makes a physicist to you?
\item[Larry:] Well first off, you have to be involved in physics somehow, I don't think you necessarily have to be, like, doing research actively to be a physicist, I just think you have to have an appreciation for physics and be involved with it in some capacity.
\end{description}
Again there is an emphasis on engaging in some activity with physics but it does not have to be research. These students also believe that you do not have to obtain an amount of knowledge to become a physicist you just have to be involved in physics in some capacity and have an appreciation for physics. Another student describes himself as an ``aspiring physicist'' but unlike students in the other categories when asked how one gets from being aspiring to just being a physicist, he replies:
\begin{description}
\item[Toby:] I really think its one of those things that you don't stop aspiring to know more, it's just human curiosity, so I don't know if you ever really stop being an aspiring physicist.
\end{description}

These students share the same belief that a passion or inherent interest in the subject is enough to be considered a physicist. The general theme of this category is that the only requirement to be a physicist is to have an appreciation or interest in physics.

\subsection{Low research/Commitment}

This category of description is unusually specific with a focus on physicists being people who are committed/serious about the subject. This category could be inclusive of students from several other categories of description as one would imagine that they are committed to physics in order to make it to the point where they are in upper-division physics classes. The distinction here, like the \textit{Low research/Mindset} category, is that the students in this category indicate no relationship to research, and its importance to them becoming a physicist. 

Both Sally and Ryan are double majors who are doing research in a physics context, and who only recently added the physics major. Both students do not point to research as being an important part of becoming a physicist. Prior identification with their other major might explain why their idea of how you become a physicist is as simplistic as deciding to pursue physics as a course of study. To students in this category, the majority of students in the other categories are physicists merely because they have declared physics as their major.

Sally describes ``declar[ing] a commitment'':
\begin{description}
\item[Int:] What makes someone a physicist?
\item[Sally:] I think they are a physicist when they have declared a commitment to it, em to the subject, whether that is declaring a major or spending time studying it\dots but making a definite commitment to the subject.
\end{description}
Ryan uses different phrasing to make the same point:
\begin{description}
\item[Int:] Do you consider the people, cause you said people who are doing a degree in physics so do you consider them physicists?
\item[Ryan:] Em, yeah if I know that they are serious about what they are doing, which most people who are into the physics department are, at least if they last a couple of semesters
\end{description}

In this category of description, a person is classified as a physicist if s/he has declared it as a major or spent a good percentage of their time studying the subject. In this paper, we examine students' perceptions of the minimal requirements to be a physicist, so commitment by itself as a category of description is acceptable. 

This perception may be a result of these students struggling in their journey to becoming a physics major:
\begin{description}
\item[Sally:] I think, I think my closest friends would understand because they have seen me struggle wanting so badly to declare a physics major but just not having the time in my schedule and they see how much I love it and so they know how dedicated I am to it.
\end{description}

However, this struggle is not a defining feature of this category. It may have informed their perception but is not a feature of their perception. In the previous extract, Sally has been struggling to declare a physics major and was asked if there were a group of people she would feel comfortable calling herself a physicist around. What is interesting is that Sally would meet almost all of the requirements indicated by any of the previous categories to consider oneself a physicist. At this advanced stage in her undergraduate development, she can't consider herself a physicist because she has not declared a physics major. It is probably this restriction that informs her perception that one must make a commitment to a subject to become a member of that community.

\subsection{Low research/Deep Understanding}

The \textit{Low research/Deep Understanding} category very similar to the \textit{High research/Deep Understanding} category, but without the emphasis on research. Both Donna and Danny, who occupy this category, are physics majors. Although they speak of an affinity for physics, they also have a strong affinity for their other subject. 

There is, however, a change in the way these students talk about obtaining  understanding/knowledge in physics.
\begin{description}
\item[Int:] Okay, do you consider yourself a physicist at the minute?
\item[Donna:] No [laughs] not by a long shot. I definitely know that, as much as I love physics, sometimes its a struggle for me. Like the other day, which was we were working on time dilation, and it wasn't clicking. And I know that takes time, and I think about things and last night all of a sudden it clicked. I don't know what I was doing but all of a sudden it made sense, oh my gosh, this is really easy, I never thought of it this way. Um, so I know I'm not a physicist yet but also I don't feel like an adult, and I'm 20 years old, so I guess it comes with time.
\end{description}
Donna is pointing to a time where a deficit in her knowledge was made apparent to her. Although Donna was able to overcome this deficit, it reinforces her perception that she could not consider herself as a physicist yet because this type of occurrence could happen again. 

With a similar efficacy theme, Danny is talking about how one becomes a physicist:
\begin{description}
\item[Int:]How do you become a physicist?
\item[Danny:]Danny: Eh, I don't know, I think for me it would it would be a confidence type of thing, being able to trust your own judgement and like how to evaluate a problem, go about problem solving, I don't know, it's hard to, when you thoroughly know the basis of physics and can apply that. 'Cause it seems once you get through [Physics] 1 and 2 and 3, you should have a solid understanding and being able to apply the fundamentals to a higher, not necessarily higher but more advanced things. For example I am doing protein nucleation with Dr. Smith\footnote{This is also a pseudonym.} and I'm applying some of the fundamentals from electrostatic fields and everything and how that works. So I don't know, just breaking it down and trusting yourself and trusting the physics.
\end{description}
Danny's reference to confidence harkens to the \textit{High research / Doing Independent Research} students' indication that they needed to develop their research to consider themselves physicists. Danny similarly wants to become self-reliant and trust in his understanding of the physics he learned before he will consider himself a physicist. 

In the \textit{High research / Deep Understanding} category, Will and Rick emphasized obtaining a certain amount of knowledge and understanding so that they could communicate/teach to others the physics they know in order to consider themselves physicists while also indicating research was a part of that process. In contrast, Danny and Donna want to obtain a certain amount of knowledge and understanding so that they can trust in themselves and their understanding of physics.

\section{Categories of Perception and Communities of Practice\label{sec:catcop}}

As we examine the categories of perception from the perspective of the communities of practice framework, we can argue that the perceptions are indicators of different stages of students' trajectories inward from peripheral to central participation. This different stage model is true for both the community of practicing physicists and the physics undergraduate community of practice. Students in the low research categories have not passed the threshold from more internal membership of the physics undergraduate community of practice to more internal membership of the community of practicing physicists. It is possible the students in the \textit{Low research/Deep Understanding} and \textit{Low research/Mindset} categories do not have an awareness that they are only on the path to central membership in the physics undergraduate community of practice and not the community of practicing physicists. Larry would seem to fit that description as he indicates a distaste for the more authentic physics experience that was Advanced Laboratory while also not pursuing an undergraduate research experience, both of which are  encouraged and available because of extracurricular activities. However, this could just be a manifestation of the stage of his undergraduate career: not engaging in more central practices of the community of practicing physicists is somewhat expected for some of the time you spend as an undergraduate physicist. Another possibility is that for some of these students it may in fact be a conscious choice by them to not pursue more central practices of the community of practicing physicists that are available to them. Whichever scenario is correct, the problem remains that students' classroom activities (although essential to becoming a physicist) are often not authentic enough to be considered central practices of the community of practicing physicists.

The \textit{Low research/Commitment} category is special because it involves students who are doing research in physics, but do not consider that as important to becoming a physicist. Instead, the act of making a commitment to the subject itself is important to them. We can explain the uniqueness of this category amongst the low research categories by the path these students took to being engaged in more central practices of the community of practicing physicists. Essentially these students bypassed the physics undergraduate community of practice because they had not made a commitment to a subject-specific identity in physics. This bypass resulted in an imposter syndrome occurring for these two students, because they felt they couldn't be physicists until they had made a commitment to a subject-specific identity as a physicist. We might find if we were to interview these students again at a later date that they would move towards an idea of a physicist as an independent researcher who conducts novel research. However, currently they seek acceptance in the community of practicing physicists by becoming a member of the physics undergraduate community of practice.

The high research categories are also interesting to look at from communities of practice perspective. It would seem a  relationship is occurring between the threshold you perceive and the legitimate practices you actually engage in. So the true threshold between the two previously mentioned communities of practice might be actually perceiving the legitimate peripheral practices of the community of practicing physicists. Essentially, the transition past the threshold occurs when you engage in practices that are open to undergraduate physicists but also extend beyond the boundaries of this community and into the community of practicing physicists. Amongst this group of students in this context, the legitimate practice that meets this requirement is undergraduate research. 

However, the \textit{High research / Doing Independent Research} category suggests that the center of the community of practicing physicists transforms once you experience the boundary breaking practice in a more authentic away. By this we mean that doing research is no longer sufficient; one must do \textit{independent} research to ``count'' as a physicist. As members of the community of practicing physicists become more aware of what the authentic practices of that community entail, their perception of the threshold to becoming a central member shifts. This shift is also true for undergraduate physics students but the shift continues to change for them as they move along their trajectory towards the center of the community. For example, you become a tenure-track professor who conducts quality novel research and central membership shifts to also include mentoring junior researchers. Perceptions of what a community's central practices entail shift as people participate in those practices.

\section{Change in perception over time\label{sec:change}}

\begin{figure*}[tb]
\begin{center}
\includegraphics[width=18cm]{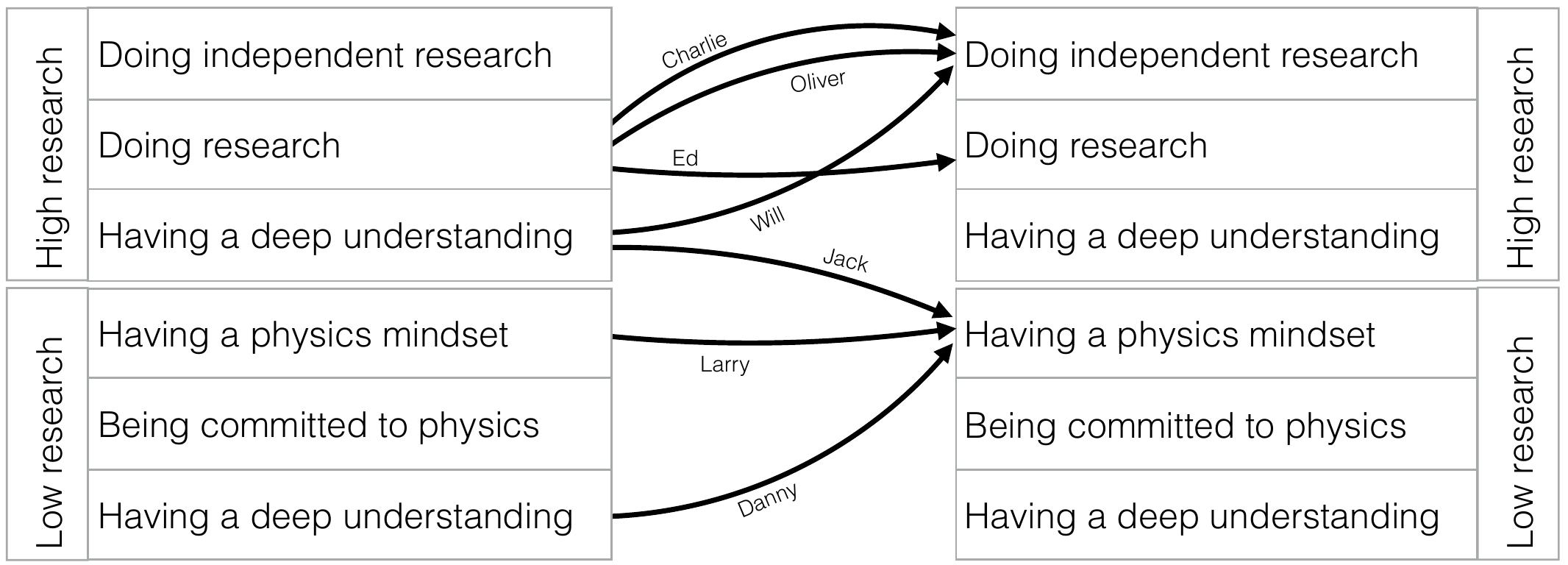}
\caption{Transitions of students between categories.}
\label{fig:transitions}
\end{center}
\end{figure*}

After we had conducted the second set of interviews, the students were placed into the categories of perception that they now occupied. We based the distribution of the students on their responses to the same phenomenographic interview but with some added questions that were aimed at reflecting on new experiences they may have had since the last interview. Five of the seven students reinterviewed made transitions from one perception to another. Two of the seven students reinterviewed did not complete transitions but did develop within the category of perception they had previously occupied (Figure \ref{fig:transitions}). In the next two sections ``Transitioning Students'' and ``Stable Students'' we examine how the students' perceptions of being a physicist changed and comment on how these transitions or lack of transition may have occurred.

\subsection{Transitioning Students}

The following students demonstrated a transition from one perception of being a physicist to another. In the case of Jack transcript was not used previously to demonstrate his initial perception and so reflections on his 1st interview are included in order to fully demonstrate the transition. The first transition we will observe is the transition to High Research/Doing Research Independence.

\subsubsection{Transitions to High research / Doing Independent Research}

The students in this section have transitioned from either the other high research categories to \textit{High research / Doing Independent Research}. Charlie and Oliver had already placed an emphasis on research but had not yet been involved in research in any fashion. Will, on the other hand, had emphasized research but said that he would feel like a physicist if he obtained an amount of understanding/knowledge so that he felt like he had mastered the subject. In the time between the first and second interviews, Charlie, Will, and Oliver (all physics majors) have undergraduate research experiences. Their perceptions have since changed.  Now Will's emphasis is much more centered on research. In the following extract, Will has altered his perspective on the obtainment of knowledge as an indicator of being a physicist to the importance of research to being a good physicist.

\begin{description}
\item[Will:] To be good at physics you need to do research, so then you have more opportunities to do more research\dots to contribute new knowledge, to discover something and share it and have people say yeah, I think you really did it.
\end{description}

In Will's second interview, he does not make references to the mastery of physics knowledge but we cannot dismiss that there might be some underlying knowledge acquisition still at play within Will. Research to fuel more research means obtaining more knowledge. However, his emphasis on discovering something ``new'', a theme of the High Research/Doing Research category, indicates a shift in perception. 

This extract also highlights a more evolved perception of a physicist as a researcher being dependent on the community they practice in for acceptance. This was absent from students in the \textit{High research/Doing Research} category, who spoke more of discovering something novel (as opposed to discovering something novel that you contribute to your respective community and being validated by that community). Just doing research is not a benchmark of being a physicist in this category. Instead, you must do research within a community and the community must value your research for you to belong to the community. 

Will's experiences of being a researcher have changed his perceptions of knowledge from a certain amount that has to be obtained to something that has to be discovered and shared with the community of practicing physicists. Before this experience research was something of an unknown quantity that had to be ticked off on the way to becoming a physicist. 
Will's research experience also indicated the importance of independence while doing research:

\begin{description}
\item[Will:] The constant directing yourself, like choosing the problem to solve, I'm learning that is a whole thing by itself, choosing the right problem to solve.
\end{description}

Oliver and Charlie demonstrate a similar transition as Will except they are moving from \textit{High research / Doing Research} to \textit{High research / Doing Independent Research}. Oliver and Charlie had previously emphasized the importance of research without expanding on why it is an important part of being a physicist. 

After a research experience between the two interviews, Oliver describes the reason doing research is an important part of being a physicist:
\begin{description}
\item[Oliver:] If you want to do physics as a career then you have to be involved in research and perpetuating the field in some manner.
\end{description}

Oliver no longer dictates an emphasis on just doing research; he qualifies it so that it has to add to the physics community in some manner. Like Will, he gets the sense that research is important so that you can become a physicist.  He also believes that it is important to the life of the field that physicists do research, because new research perpetuates the field. Oliver now has a sense of responsibility to his community that it is important for him, as a future professional physicist, to further his field of study. 

Charlie seems to have come to a similar conclusion about the purpose of research:
\begin{description}
\item[Charlie:] I guess my idea of a physicist is just much different than knowing physics knowledge..I guess its what you do with the knowledge and your specifically using the knowledge to advance physics, specifically using it for physics, doing research now.
\end{description}

Also like Will, Oliver and Charlie place an emphasis on the need for independence in one's research in order to transition from being a student to being a physicist:
\begin{description}
\item[Oliver:] I think when their research becomes more independent, I think that's a good distinction point, grad school is basically to learn how to do your own experiments, as soon as you stop working for your mentor and you're branching off into your own direction, I think that's maybe the point (you become a physicist)
\end{description}

Charlie's research experiences since the last interview have made him realize that doing research can mean many things and to be a physicist is more encompassing than just actually participating in the process. Instead, it is making decisions on that process and direction, independent of a supervisor.

\begin{description}
\item[Int:] So there is an aspect of independence to it then?
\item[Charlie:] Yeah, if you can do it on your own, like not entirely, no physicist is going to do it entirely on their own, if you don't have to go back and ask what do I do now, or how do I do that then that determines you as more of a physicist.
\end{description}

All three students demonstrate an ownership of research and make explicit reference to independence as being an important element of research being an indicator of being a physicist. The transition these students demonstrate is understandable from the perspective of the experiences that they have engaged in between interviews. These students have engaged in practices that expose them to the realities of research in physics. An expected norm of the purpose of research in physics is that the researcher is attempting to further their communities understanding of the concept/phenomenon being investigated. This prevalent norm in research groups may have been a significant influence on these students perceptions of the purpose of doing research and what it means to a physicist. These students have probably also had experiences of research that is tightly controlled while also noticing that professors, post-docs and graduate students operate with much more independence. 

\begin{description}
\item[Oliver:] But when you are just sitting in front of it for hours while it just scanned, that's not something I want to do and I know that is not what my professor does or my post-doc does but I don't want to spend four years doing that and then telling someone else to do that.
\end{description}
As they are involved more and more in research, their perceptions of research will likely continue to evolve.

\subsubsection{Transitions to Low research/Mindset}

Danny is a double major in physics and engineering. Jack is in the process of declaring second major in engineering (in addition to his physics major). Both of these students have had research experiences in physics between the first and second interviews. Danny and Jack are in the process of transitioning from perceiving that one becomes a physicist by acquiring a certain amount of knowledge (\textit{Deep Understanding}) to perceiving that being a physicist is a mindset (\textit{Low research/Mindset}). It is unclear as to whether these two students have fully transitioned to this new perception. Both Danny and Jack in their interviews reflect on past views while arguing new perceptions.

In Jack's first interview, his perceptions of when one becomes a physicist are centered on the acquiring of a qualification and the obtainment of a certain amount of knowledge. In the following extract, he reflects (accurately) on his previous interview and his previous perceptions:
\begin{description}
\item[Jack:] I'm pretty sure my answer was a physicist is someone who gets like a degree, goes for the Ph.D. or masters in physics something like that\dots and I still have the achievement type idea, I try and shy away, but I still have that thought in my head\dots and a lot of its content knowledge, but it's not a necessity to being a physicist.
\end{description}

This extract indicates a person who is currently in conflict with his previous perceptions of when one becomes a physicist. He reflects on his previous thoughts while assessing that they may not conform to the way that he is thinking now. We note that Jack uses educational jargon such as ``content knowledge.'' The initial part of the interview is an introduction to why the students are being interviewed, but these terms are never used by the researcher to describe the study. Jack's research experience between interviews was a Research Experiences for Undergraduates (REU) summer research position. As part of this position, the cohort of summer students presented their research weekly to each other.  Part of this cohort were three physics education research students. It is likely that Jack picked up on terms such as ``content knowledge'' from these talks and interactions with these students. 

In the same interview he describes being a physicist as being a certain way of thinking and emphasizes the physicist as the problem solver:
\begin{description}
\item[Jack:] Honestly what I view the physics major has really given me, it's given me a lot of things but the most important thing is, really, really, really gotten good at problem solving, kind of what I look at as a physicist now, is he is a master problem solver.
\end{description}

Later in the interview, Jack again touches on this idea of being a physicist as being a way of thinking when describing the effect his research experience had on him:

\begin{description}
\item[Jack:] So my idea of a physicist before hand was just academics who do research and teach a class, they do hardcore research or something like that, but sitting through the meetings it also made me realize that it was defending physics and criticizing other peoples work to make sure it is actually what it is…they want you to think in a certain way
\end{description}

Jack seems to be either merging or moving away from the idea of the physicist as a reseacher to the perception that a physicist has a certain mindset. Both Jack and Danny do not dismiss research as being unimportant, but they do underplay its importance in being a physicist.

We placed Danny after his 1st interview into Low Research/Deep Understanding.  In his second interview he continues to emphasize the need for understanding the material he deals with in class. 

He continues to lower the emphasis on the need for research without dismissing it completely:
\begin{description}
\item[Danny:] I don't think you need a Ph.D. to be a physicist I think you need to put a lot of time into it, equivalent to maybe a Ph.D\dots I think the biggest thing separating, someone who is motivated to learn and (someone) who is a physicist is experience and intuition, your intuition about problems and also being up to date on you know what is real and what is the current research at, I think that's part of being a physicist as well.
\end{description}

Danny's thoughts about intuition and its role in problem-solving would seem to indicate a shift towards the idea of physics as being a way of thinking and approaching problem-solving. For both Danny and Jack, these transitions away from an emphasis on research and obtaining a certain amount of knowledge are perhaps not surprising. The dual nature of their studies means that they both have competing subject specific identities and a negotiation process over which professional identity they will settle on. They indicate in the second interview that they wanted a future profession that blended the worlds of physics and engineering together. In fact, their two subject specific identities and potential professional identities are often talked about in concert with each other. Jack indicates that he began exploring engineering as a fallback option because he began to worry about the amount of time needed to be devoted to becoming a physicist.

\begin{description}
\item[Jack:] I'm not sure if that's what I want to do, that's a huge commitment and if it didn't go right I don't know what would happen, so that's when I started taking electrical engineering classes, maybe I could double major, that way I could get a job right after undergraduate school.
\end{description}

Danny also talks at length about how physics is what he is more interested in, but that engineering provides a useful fallback for him. 

\begin{description}
\item[Danny:] My idea was physics is what I am interested in and I just wanted to study it for the sake of knowing, and mechanical engineering, I was thinking I was just going to get a job in my fall back.
\end{description}

However, at no point does Danny or Jack talk about engineering in a negative way and instead focus on the positives of each subject and what a professional identity in either would involve. In fact for both of these students a combined professional identity of an engineering physicist is ideal for them:

\begin{description}
\item[Jack:] I have decided if I could mix what I did with Dr. Black, with my engineering Dr. Plum experiences I would be peak happiness because I loved what I did with him and I really enjoy what I am doing with Dr. Plum, engineering studies, I really love that stuff, so if I could somehow combine them, I feel like I would be right where I am supposed to be.
\end{description}
Danny and Jack's movement towards a perception of a physicist as a mindset might be them trying to identify the aspects of their physics identity they can blend with aspects of their engineering identity. This blending is so they can develop a professional identity that incorporates the two identities.

\subsection{Stable Students}

Of the seven students reinterviewed, two of these students demonstrated relative stability in their perception of being a physicist. We placed Ed in the \textit{High research / Doing Research} category after his first interview, and we placed Larry in the \textit{Low research/Mindset} category after his first interview. Between the first and second interviews, Ed (a double major in physics and chemistry) continued with a physics research experience. Larry (a physics major) did not have any research experience between the two interviews but did take the Advanced Laboratory course. 

In the following section, we examine how Larry and Ed maintain their original perception of being a physicist with relative consistency. Ed continues to focus on research as the indicator of becoming a physicist and maintains that research is important because of its ability to answer unanswered questions. He highlights the goal of the obtainment of a publication as an indicator of being a physicist:
\begin{description}
\item[Ed:] I'm pretty sure I said this last time but a publication is a benchmark, once you're published, if you're not a physicist then you are pretty close to becoming one. 
\end{description}

As indicated in the quote, this was his perception during the previous interview, and this has not changed in the period between the two interviews. What has changed in his perception is how the answers to the unanswered questions realistically come about: 

\begin{description}
\item[Int:] Has it (the research experience) changed at all the way you view physics? 
\item[Ed:] Definitely, as I mentioned we wait 5 hours for things to cool up, cool down, and when you factor that it in, I suppose its not as glamorous than when you are coming right in and saying, wow they are making all of these discoveries every single day.
\end{description}

Ed now has an experience-based perception of what it means to be a physicist. Although this experience has not altered his perception of when he will become a physicist, it has altered his perception of how the discoveries that physicists make actually occur. Ed's assertion that the publication is the threshold he would judge his transition into being a physicist is unfortunately not explored in any great detail in the interview. The publication as threshold is interesting as Ed could view it as a rite of passage of acceptance within the community or the point at which he has contributed new knowledge to the physics community. If he perceived this publication as being one in which he was the first author, he could also have perceived this threshold as being his proof of independent research. The researchers decided to keep Ed in High Research/Doing Research category because he does not speak about the importance of independence in his research. He also indicates that he is just starting out this process of learning about research and how to conduct it. 

\begin{description}
\item[Ed:] I am developing lab techniques, I am using a UV-VIS spectrometer in my current research. I do data analysis and things like that\dots not that I have gotten a lot out of it yet because it takes forever.
\end{description}

Not all research experiences are the same and Ed is at the beginning of his and so might not have been introduced to the independence aspect of research as of yet. More time with his research group and conducting research may eventually result in a transition from one perception to another. Larry, on the other hand, is probably the most stable student between the 1st set of interviews and the 2nd set. He is persistent in his perception that he is a physicist at this point in his academic career and that being a physicist is viewing the world in a particular way or having a particular mindset.

\begin{description}
\item[Larry:] It's sort of like a mindset (being a physicist) where you have a difficult situation\dots then establish what you know and work those assumptions\dots try to analytically figure out a solution.
\end{description}

As with a lot of students who believe that being a physicist is a mindset, Larry alludes to this mindset being an approach to problem-solving that is unique to a physicist. Larry maintains that this mindset of problem-solving often informs that person's view of the world. Unlike Ed, who also display a consistency with his category of perception, Larry's particular perception does not alter to incorporate new types of physics experiences because he has not had a new type of physics experience. This time around though, Larry's view about research has slightly changed:
 
\begin{description}
\item[Larry:] Yeah I think so, it doesn't hurt your case if you're doing research, but I feel like, em, sort of like as long as you are synthesizing knowledge em then that's like the main criteria.
\end{description}

Larry has come around to the idea that it couldn't hurt to do research but is still persistent that it is not necessary to be a physicist. At this stage in Larry's undergraduate career, he has seen his peers seek out research experiences and so is not blind to the fact that it is something that is done. 

Although Ed did not explicitly demonstrate a full transition from one category to another, he did exhibit small but important shifts in his descriptions of what it means to be a physicist. These revised descriptions inform a complete description of their particular perception of when one becomes a physicist. Ed now understands that research is more than just something you do; it can be tedious and involve monotonous activities, but it can also be exciting when analysis actually reveals something new. The opposite to this would be Larry who has not gained a research experience nor has he really been exposed to a physics experience that could be considered drastically new. As a result, his perception has remained fairly consistent and unchanged. Previously it was indicated that Danny and Jack may be transitioning from one category to another because they are double majors. The argument being that they are trying to maintain an identity as a physicist even if their trajectory does not take them along a traditional physicist route. The same reasoning may apply to Larry, who continues to maintain that he does not want to go to graduate school and instead is considering becoming a teacher. Larry's perception of being a physicist is very accommodating for those studying physics who wish to enter into the teaching profession.

\section{Transitioning Students and Communities of Practice\label{sec:changecop}}

If we consider the practice of research as an a priori legitimate peripheral practice of the community of practicing physicists then it appears participating in this practice changes your perception of what central membership of the community of practicing physicists appears to be. Evidence of this claim is in the observed movement of Charlie, Oliver, and Will toward research that is self-guided after having engaged in undergraduate research between interviews.

Jack and Danny's movement, on the other hand, is more difficult to describe with the communities of practice framework. We propose that because Jack and Danny intend on pursuing a professional identity in engineering or a blend of engineering and physics that they do not perceive research as a legitimate peripheral practice for themselves. They are on trajectories to central membership in another professional identity but do not want to reverse the progress they have made in the community of practicing physicists. To hold this progress, they are focused on a perception of physics being a mindset which can be applied to their new professional identity. 

Larry's case is interesting, and it would be fruitful to investigate further students who have persisted with his perception while, not exploring new experiences in physics. As argued previously for his category of perception, students with the mindset perception may be unaware of the need to engage in legitimate practices outside of those offered in the classes they take as undergraduate physicists. It is still possible that Larry will become a central member of the physics undergraduate community of practice by meeting the requirements of a physics degree. It is also possible he will not, given that the majority of the physics majors have completed some form of undergraduate research. By not engaging in the buffer-breaking practice of research, he may not become a central member of the community of practicing physicists. Maybe this is not his intention: Larry has indicated before that he is interested in pursuing high school teaching as a possible professional identity. A physics teacher identity is another identity that will incorporate legitimate peripheral practices of the community of practicing physicists. However, there will also be a buffer to becoming a central member of this community. It is also possible that Larry is aware of the distinctions between the communitiy of practicing physicists and the undergraduate community of physicists and is choosing to disregard those differences in favor of a more inclusive community of physics thinkers. A further interview with Larry could explore this in greater depth and indicates the need to further explore the journeys of students with this perception of being a physicist.

\section{Discussion\label{sec:discussion}}

The majority of students interviewed in this research study fell into research emphasized categories. The fact that research features so prominently in the descriptions of being a physicist by these students is a clear indication that faculty at KSU promote the idea of the physicist as a researcher. The high number of students in the research category emphasizes research as being an important part of being a physicist. However, students in this category often did not have a research experience and so were unsure of what it entails. A high percentage of the total students taking part in this study being initially in this category means that the idea of the physicist as the researcher is being promoted. Physics faculty need to inform their students of what research typically entails, how it is carried out and how it relates to them being a physicist. This communication should not be the limited to the staff teaching the major orientated physics classes and should instead extend to all faculty to be more aware of how they discuss research. Perhaps deconstructing why research is important to the community of practicing physicists and explaining some of the realities of the research processes will help students develop an identity as a professional physicist quicker. Physics departments could take their cues from undergraduate research programs that often emphasize students' professional development through seminars, field trips, and discussions. The aim of these activities is to expose students to various science careers and to inform them on what research typically entails\cite{Hunter2006}. 

The previous recommendation argues for a refinement in the approach to undergraduate research but even with out these refinements undergraduate research has been demonstrated to have been an informative experience for the students interviewed. Students interviewed a second time in this study typically demonstrated a fuller understanding of the realities of conducting research in physics and its importance after having conducted undergraduate research of their own. This would indicate that an undergraduate research experience can be a pivotal experience in a student's trajectory to becoming a member of the community of practicing physicists. For the six out of the seven students who we reinterviewed it either resulted in a change in perception of what it means to be a physicist or further expansion of their current perception. Previous research has demonstrated\cite{Hunter2006} that students' gains in confidence can be attributed to their change in perception that the research they conduct as undergraduate researchers can make a useful contribution to the field. The following extract from Will would also seem to indicate that his perception of his undergraduate research experience helped him to engage in more legitimate peripheral practices of the community of practicing physicists.

\begin{description}
\item[Will:] I'm learning a lot\dots it's not like in a classroom when you're given problems and you have to figure them out. It's a real physicist's experience where you have to look up papers and try and understand some guy talking about something that you have never heard of before and trying to mine relevant information out of that. That is what you are trying to do and not really having a set path, I guess, having to make it up yourself, thinking critically about data and how to acquire it\dots interpreting things, it's really hard to make the distinction between this is good enough and this is not good enough. Like, I have scattering patterns, and I have to decide: did I actually get something this time and is it close enough to what the theory says to say, I get. It's weird, I like it more and more, the more I do it, but it's weird. It's not one of the sexy things when I thought about doing physics.
\end{description}

This quote relates to the community of practice model of identity development. When students engage in undergraduate research, they are being guided by more central members of the community of practicing physicists through engagement in the authentic practices of that community. Conducting research also effects students perceptions of what authentic practice looks like for a physicist. For several of the students who transitioned from one perception to another, this transition also resulted in them reevaluating what legitimate peripheral practice looks like when they have become more central members of the community of practicing physicists. It follows then that perceiving being a physicist and the subject of physics in different ways might encourage students to engage in different practices or engage in the practices that they currently engage in a more authentic way. This change in perception could enable a trajectory to a more central membership. We need to reinforce this claim by conducting further research on more central persons within the community of practicing physicists to ascertain what perceiving being a physicist at this point in their professional career and identity might look like.

There is an underlying assumption being made in the previous discussion about becoming a member of the community of practicing physicists: all upper-division undergraduate physics students intend to become a member of this community. Several students in this study through either their initial perceptions or the transitions in perceptions they undergo would indicate that this is a false assumption. The question then is whether we are forcing out potential physicists because, as an undergraduate, they do not conform to the expectations of the central members of the community of practicing physicists. Students who occupied low research categories tended to be either physics majors with a double major or physics majors who do not intend on going to graduate school or becoming a physics researcher. 

For example, Toby and Larry intend on becoming physics teachers. With this career intention, these students are less likely to place importance on research especially when it comes to being a physicist. If they did place this importance on research, they then might feel that they do not belong in the community of practicing physicists and become alienated. This lack of belonging could cause discouragement and result in retention issues for these types of students. Similarly, if we look at Danny and Jack, who are in the process of transitioning from one perception to another. We could argue that their movement away from the perception that highlights research as being an important part of being a physicist is indicative of their intended career path. They hope to blend physics and engineering together for a professional identity that combines the two. Again though the feeling that they might not belong to the community of practicing physicists might be communicated to these students as a result of them rejecting the legitimate peripheral practices of that community. 

In the case of both of these groups, the question becomes whether it is necessary for these groups of students to become more central members of the community of practicing physicists. Is it sufficient for these students to be central members of the community of practicing teachers or a subject specific community? The previous question needs to be answered more thoroughly by following students like Larry and Danny into their respective careers and seeing which membership to which communities of practice become important to them. 

Talking about membership of the community of practicing physicists is complicated and troublesome due to the community being ill-defined. All of these students will join communities of practice that are physics related, and these will be sub-communities of the community of practicing physicists. These students will still be a member of the community of practicing physicists just perhaps not central. At this point in their undergraduate career, this may not be clear to students such as Danny or Larry. The emphasis on research may, in fact, be a negative norm\cite{DallAlba2006} for students like Danny and Larry. They may feel as if they are failing to align with the norms of the overarching community of practice and for this reason no longer belong. The question here would be whether the community of practicing physicists want these individuals as members. If we do, then we should support these students and the respective sub-community of practice. We should help students so that they can align their intended career goal with related authentic practices of a physicist to ensure they feel welcome in the community of practicing physicists. 

We have identified the undergraduate research experience as being a legitimate peripheral practice of both communities of practice we have investigated. Experience with this legitimate practice breaks through the barrier between the community of practicing physicists and the physics undergraduate community of practice. However, placement into summer research positions can be quite competitive and undergraduate research within an institution can be limited geographically and in how authentic it can be. The central members of the community of practicing physicists should look to merge these two related communities more effectively. One way of doing this is by changing the legitimate peripheral practices of the physics undergraduate community of practice. Components of course programs should highlight various aspects of authentic practice and integrate these practices into students' coursework\cite{DallAlba2006}. 

The transitions students took from one category to another, and the movement of students towards perceiving the importance of independent research indicate that time is a factor on the completeness of the categories identified. The effect of time makes sense when dealing with a research methodology that focuses on experiences. As students progress with their undergraduate studies, they will invariably have new physics related experiences that inform their perceptions of being a physicist. Some of these experiences will be more pivotal than others such as an undergraduate research experience compared to a homework problem for their electricity and magnetism class. In future research, we should aim to discover how these perceptions continue to change over time. We should also aim to identify other particular physics related experiences that can have a greater effect on student's trajectory like undergraduate research does, to becoming a member of the community of practicing physicists.

This leads to a final point on the methodology. The phenomenographic research methodology is focused on distilling the critical aspects of particular experiences, and is therefore more broadly applicable in the Physics Education Research community (e.g. \cite{Walsh1993,Ingerman2009Physics,Neuman1997,Lo2014,Irving2013,Irving2015csse,irving:198,Madsen2015,Madsen2014a}). It provides the opportunity to explore students' and faculty's experiences at a deeper level and discover the nuances between their experiences.  This process leads to the development of new and exciting research questions while also providing the leverage for proposing hypotheses for researchers' observations. Whether exploring students' experiences of the concept of energy or exploring their positive affective experiences in a laboratory, the phenomenographic research methodology promotes a generative understanding of these phenomena. 

\section{Conclusion\label{sec:conclusion}}

This research presented six categories of perceptions of being a physicist. Three of the categories focused on research as being a key part of students' perception of being a physicist and three focused on more diverse perceptions of physicists. Over time, students transition between perceptions. We identify undergraduate research as an important threshold experience that encourages transitions into the community of practicing physicists.

\section{Acknowledgments}

This work was partially supported by the KSU and MSU Departments of Physics.  Any opinions, findings, and conclusions or recommendations expressed in this material are those of the authors and do not necessarily reflect the views of the departments.


\end{document}